\documentclass[preprint,showpacs,apl,preprintnumbers,superscriptaddress,amsmath,amssymb]{revtex4-1}
\usepackage{graphicx,dcolumn,bm,color}

\begin{document}

\title{Localization effects in the tunnel barriers of phosphorus-doped silicon quantum dots}

\author{T~Ferrus}
\email{taf25@cam.ac.uk}
\affiliation{Hitachi Cambridge Laboratory, J~J~Thomson avenue, CB3 0HE, Cambridge, United Kingdom}

\author{A~Rossi}
\affiliation{Hitachi Cambridge Laboratory, J~J~Thomson avenue, CB3 0HE, Cambridge, United Kingdom}

\author{W~Lin}
\affiliation{Quantum Nanoelectronics Research Centre, Tokyo Institute of Technology, 2-12-1 Ookayama, Meguro-ku, Tokyo, 152-8552 Japan}

\author{D~A~Williams}
\affiliation{Hitachi Cambridge Laboratory, J~J~Thomson avenue, CB3 0HE, Cambridge, United Kingdom}

\author{T~Kodera}
\affiliation{Quantum Nanoelectronics Research Centre, Tokyo Institute of Technology, 2-12-1 Ookayama, Meguro-ku, Tokyo, 152-8552 Japan}
\affiliation{Institute for Nano Quantum Information Electronics, University of Tokyo, 4-6-1, Komaba, Meguro, Tokyo, Japan}
\affiliation{PRESTO, Japan Science and Technology Agency (JST), Kawaguchi, Saitama 332-0012, Japan}

\author{S~Oda}
\affiliation{Quantum Nanoelectronics Research Centre, Tokyo Institute of Technology, 2-12-1 Ookayama, Meguro-ku, Tokyo, 152-8552 Japan}

\begin{abstract}

We have observed a negative differential conductance with singular gate and source-drain bias dependences in a phosphorus-doped silicon quantum dot. Its origin is discussed within the framework of weak localization. By measuring the current-voltage characteristics at different temperatures as well as simulating the tunneling rates dependences on energy, we demonstrate that the presence of shallow energy defects together with an enhancement of localization satisfactory explain our observations. Effects observed in magnetic fields are also discussed.

\end{abstract}

\pacs{72.80.Cw, 73.63.Kv, 73.23.Hk, 73.20.Fz, 73.40.Gk, 71.55.Cn}
\keywords{silicon, quantum dot, Coulomb blockade, localization, cotunneling, defects, negative differential conductance}
                        
\maketitle
  
\section{Introduction}

The consequences of localization and disorder on the transport properties, in particular, on the mobility of devices have been investigated extensively in bulk and in two-dimensional semiconductors \cite{Anderson, Davies}. One of the most noticeable and known effect is hopping conduction where electrons are allowed to hop between impurities sites, leading to a decrease in the material conductivity with temperature. This was first theoretically described by Abrahams \cite{Abrahams} to explain the transport properties of random media. In this case, the boundary between ballistic and diffusive transport is well represented spatially by an ellipsoid which major and minor axis are given respectively by the hopping and localization lengths. Unless screened by a metal gate or short-range disorder, electron-electron interaction is often competing with localization, leading to a modification of the conductivity at low temperature \cite{Ferrus1} and eventually to a metal to insulator transition \cite{MIT}. In nanostructures and, in particular in quantum dots, the localized region is defined by the device geometry, principally the number, the shape, the nature and the orientation of interfaces. Due to dielectric screening, the effective Bohr radius at trapping sites at the edges of the structure can significantly increase and electron localization becomes possible. Such phenomenon have been already observed at the periphery of dots fabricated within an AlGaAs-GaAs heterostructure \cite{Zhitenev} and Coulomb blockade has been found to be affected via a change in the dot capacitance.

In doped devices, like phosphorus-doped silicon single electron transistors, the localized region has a more profound effect owing to the presence of dopants at the edge of the structure and so, in close proximity to the Si-SiO$_2$ interface \cite{Ferrus2}. In presence of Coulomb interaction, charge rearrangement is possible and becomes an effective mean of reducing the charging energy. It may also be responsible for the observed reduction in the effective dot dimension due to electron trapping at the edge of the structure and the subsequent change in the electrostatic potential. 

Despite this, the outer region of the dot is not the only place where localization has a non-negligible influence. Randomness in dopant distribution may lead to traps being present at or close to the tunnel barrier. As a result, the transmission probability, and so, the device conductivity, is expected to be significantly affected. The situation where a single dopant is unintentionally located at the barrier has already been investigated \cite{Sanquer, Sanquer2} in silicon nanostructures. However, most studies report on resonant features following the alignment of the quantum dot levels with the D$_0$ or D$_-$ states of the impurity \cite{Rogge}, a system that drives interest as it could be applied to spin filtering \cite{spin blockade} in spin qubit architectures \cite{spin qubit}. 

In this article, we have used a highly doped silicon single electron transistor (SET) and show that an additional localization process may lie at the tunnel barrier. Caused by natural defects or more generally, by an asymmetry in the barrier shape, the effect can enhance significantly the electron trapping mechanism. In the first section, we discuss the observation of a negative differential conductance as well as its dependence on gate voltage. We then describe the localization process through the temperature dependence of the conductivity and simulations. The analysis of experimental results allows distinguishing between static and dynamic effects in the tunnel barriers. In a second paragraph, we describe and interpret the magnetic field and temperature dependences in the low temperature regime before conclusions are drawn on the origin and potential implication of this effect.

\section{Device and experimental setup}

Devices were fabricated from a silicon-on-insulator (SOI) substrate with a 40 nm-thick silicon layer that was subsequently implanted with phosphorus to obtain a dopant density of 3$\times 10^{19}$ cm$^{-3}$ after thermal anneal. At such a doping density, the largest areas of the contact leads remain metallic whereas high density defect surfaces partially deplete the nanostructures and locally increase the localization at its edges \cite{Ferrus2}. The quantum dot and its controlled in-plane gate $V_{\textup{\tiny{g}}}$  were defined by electron beam lithography during a single step process. The dot diameter was then reduced to about 120 nm after thermal oxidation, leading to the formation of a 20 nm oxide layer (Inset Fig. 1). All measurements were taken in an Oxford Instrument Heliox$^{\textup{\tiny{TM}}}$ cryostat that was fitted with a 7 T magnet and whose base temperature was 300 mK. The current $I_{\textup{\tiny{SD}}}$ was measured  using a Hewlett Packard 3458A multimeter with a 10$^8$ V/A amplifier on battery. The temperature of the device was controlled by an Oxford Instrument Intelligent Temperature Controller (ITC4) from 300 mK up to 150 K using RuO$_2$ and Cernox calibrated thermometers.

\section{Experimental results}

\subsection{Negative differential conductance}

\begin{figure}
\begin{center}
\includegraphics[width=85mm, bb=2 1 241 170]{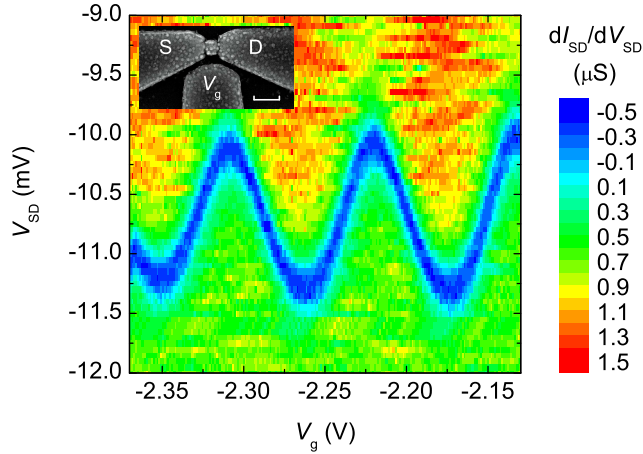}
\end{center}
\caption{\label{fig:figure1} Position of the NDC in $V_{\textup{\tiny{SD}}}$ and $V_{\textup{\tiny{g}}}$ at 300 mK. The inset shows a Scanning Electron Microscope (SEM) image of the device with source (S), drain (D) contacts as well as the SET gate $V_{\textup{g}}$. The bar scale represents 200 nm.}
\end{figure}

The device is characterized by regularly spaced Coulomb diamonds and so, it operates as a single quantum dot at low temperature. This will not be discussed here, but one may report to previous observations \cite{Rossi}. However, a negative differential conductance (NDC) can clearly be noticed in the region of negative source-drain bias $V_{\textup{\tiny{SD}}}$ (Fig. 1). Such feature could be found in single quantum dot where additional parallel channels via edge states are opened for tunneling or when trap-assisted tunneling is present \cite{Pierre} or in the case of spin dependent transport \cite{Hitachi}. Here, the position of the NDC clearly follows the shape of the Coulomb diamonds, at a fixed distance from their edges. This rather suggests that the existence of the NDC is associated with tunneling events in the quantum dot. Previous studies on the same quantum dot have shown that such a NDC was offsetting the electron tunneling detection point of a nearby SET \cite{Rossi} and that an asymmetry in the tunneling rates could explain the observed phenomenon. Still, this observation alone does not enlighten on the fundamental electronic process responsible for such asymmetry in the tunneling rates and so, additional experiments are required.

\subsection{Localization effects in the tunnel barrier}

\subsubsection{Barrier asymmetry}

The asymmetry of the tunnel barrier rates can be assessed more directly by determining the dependence of the multi-electron tunneling rates for each barrier \cite {Rates} $\it{\Gamma}_{\textup{\tiny{S, D}}}$ on $V_{\textup{\tiny{SD}}}$. Extracting separately the values for $\it{\Gamma}_{\textup{\tiny{S}}}$ and $\it{\Gamma}_{\textup{\tiny{D}}}$ from the total tunneling rate $ I_{\textup{\tiny{SD}}}/e = \it{\Gamma}_{\textup{\tiny{S}}}  \it{\Gamma}_{\textup{\tiny{D}}}  / \left( \it{\Gamma}_{\textup{\tiny{S}}} + \it{\Gamma}_{\textup{\tiny{D}}} \right)$ and the dependence of the current on temperature is generally challenging. This is due to the presence of both elastic and inelastic processes and their complex dependences on source-drain and gate biases. 

\begin{figure}
\begin{center}
\includegraphics[width=85mm, bb=2 2 224 171]{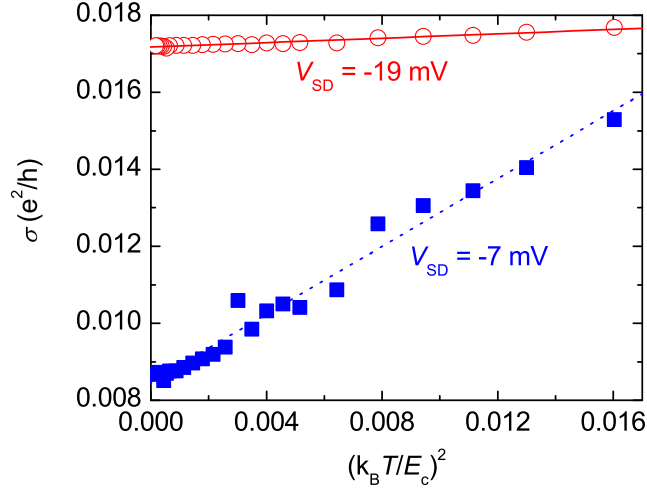}
\end{center}
\caption{\label{fig:figure2} Conductivity in the inelastic cotunneling regime at different values of $V_{\textup{\tiny{SD}}}$.}
\end{figure}

However, it is possible to estimate the product of the tunneling rates by observing the quadratic dependence of the conductivity in temperature below 3 K, a behavior that is characteristic of inelastic cotunneling \cite{Furusaki} (Fig. 2). Indeed, within this regime the conductivity is given by \cite{Glatti}:

\begin{eqnarray}\label{eqn:equation1}
\sigma_{\textup{\tiny{SD}}}  \approx  \frac{2 h e^2}{3} \it{\Gamma}_{\textup{\tiny{S}}}  \it{\Gamma}_{\textup{\tiny{D}}}  \frac{1}{{E_{\textup{\tiny{C}}}}^2{\Delta F}^2} \left[ \left( k_{\textup{\tiny{B}}}T \right)^2 + \left( \frac{eV_{\textup{\tiny{SD}}}}{2\pi} \right)^2 \right]
\end{eqnarray}

where $E_{\textup{\tiny{C}}} $ is the dot charging energy, $\Delta F = -e \alpha_{\textup{\tiny{D}}} V_{\textup{\tiny{SD}}} - e \alpha_{\textup{\tiny{g}}} V_{\textup{\tiny{g}}}$ the change in the free energy of the system on the tunneling of an electron in or out the quantum dot, $\alpha_{\textup{\tiny{D}}}$ and $\alpha_{\textup{\tiny{g}}}$, the level arms of the drain and gate respectively.

On the other hand, it is possible to estimate the tunneling rates by taking into account the tunnel probabilities of an electron through a barrier at a specific energy level $E_N= N E_{\textup{\tiny{C}}}+ E_0$ in the quantum dot. The use of the parabolic barrier approximation \cite{Rossi} allows extending the range of validity of the tunnel rates to large $V_{\textup{\tiny{SD}}}$. Following, a method similar to the one used for square barriers \cite{MacLean}, we found :

\begin{eqnarray}\label{eqn:equation2}
{\it{\Gamma}}_{\textup{\tiny{D}}} \propto  \frac{4 \Delta E}{h} \textup{exp} \left[ \alpha \left(\Delta F+ E_N\right) +\beta \right]\,f( e \alpha_{\textup{\tiny{D}}} V_{\textup{\tiny{SD}}}, T_e )
\end{eqnarray}

$f(E, T)$ is the Fermi distribution of the drain reservoir, $T_e$ is the electron temperature, $\Delta E$ the level spacing of the dot. Coefficients $\alpha$, $\beta$ are given respectively by :

\begin{eqnarray}\label{eqn:equation3a}
\alpha = -\frac{\pi}{2} W \left( \frac{2 m^* V}{\hbar^2} \right)^{1/2}
\end{eqnarray}

\begin{eqnarray}\label{eqn:equation3b}
\beta = \frac{\pi}{2} W \left( \frac{2 m^* }{\hbar^2 V} \right)^{1/2}
\end{eqnarray}

where $m^* =0.19 m_{\textup{e-}} $ is the  effective mass of silicon, $V$ and $W$ respectively the height and width of the tunnel barrier (defined at the Fermi energy of the reservoir).

\begin{figure}
\begin{center}
\includegraphics[width=85mm, bb=2 2 224 168]{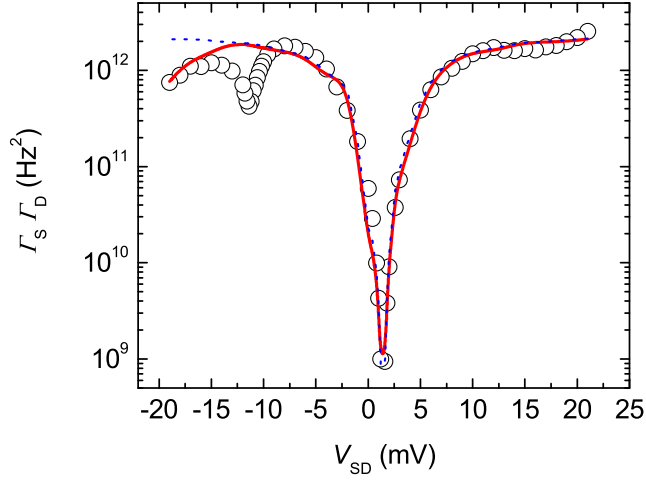}
\end{center}
\caption{\label{fig:figure3} Variation of the product of the tunneling rate of the source ($\it{\Gamma}_S$) and the drain ($\it{\Gamma}_D$) with source-drain bias and corresponding simulations for parabolic barriers (dotted lines) and in the case of energy dependent barrier width (solid line).}
\end{figure}

The value of $\it{\Gamma}_{\textup{\tiny{S}}}  \it{\Gamma}_{\textup{\tiny{D}}}$ is then calculated by summing over all possible tunneling events, including elastic and inelastic contributions for the different energy levels of the quantum dot (ground and first excited states). The behavior of $\it{\Gamma}_{\textup{\tiny{S}}}  \it{\Gamma}_{\textup{\tiny{D}}}$ is well reproduced for both positive and negative $V_{\textup{\tiny{SD}}}$ (especially at small biases) (Fig. 3) if one takes into account the dependence of the electron temperature on source-drain bias due to acoustic phonons \cite{ElecTemp}:

\begin{eqnarray}\label{eqn:equation4}
T_e \approx T_0 \left[1+ \left( \frac{e \alpha_{\textup{\tiny{D}}}}{ 3.53 k_{\textup{\tiny{B}}} T_0} \right)^{7/2} \frac{{V_{\textup{\tiny{SD}}}}^4}{{V_0}^{1/2}} \right]^{2/7}
\end{eqnarray}

where $T_0$ is the phonon bath temperature and the normalization constant $V_0 = 1$ mV.

The drain and gate level arms can be extracted from Coulomb diamonds and we find $\alpha_{\textup{\tiny{g}}} =0.048$ and $\alpha_{\textup{\tiny{SD}}}=0.44$.
From best fits, we obtain $W_{\textup{\tiny{S}}}=8.0$ nm and $W_{\textup{\tiny{D}}}=13.3$ nm  for the source and drain barrier widths and, $V_{\textup{\tiny{S}}}=5.4$ meV and $V_{\textup{\tiny{D}}}=11.8$ meV respectively for the corresponding barrier heights. The values for the barrier widths are compatible with the ones obtained from SEM observations and suggest the existence of an asymmetry in the device structure, and so, an asymmetry in the tunneling rates. It is interesting to notice that the potential barriers are small implying parallel conductive path at relatively small bias values, an effect explaining the additional conductivity background for $\mid V_{\textup{\tiny{SD}}} \mid >10$mV.

\begin{figure}
\begin{center}
\includegraphics[width=85mm, bb=2 2 222 164]{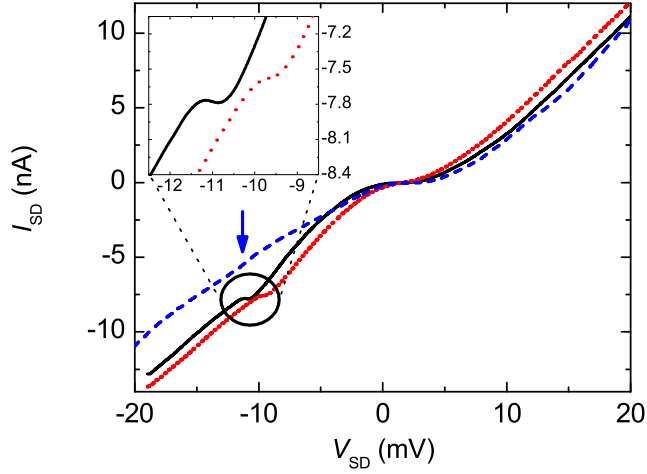}
\end{center}
\caption{\label{fig:figure4} Current-voltage characteristics before (solid line) and after thermal cycle to 150 K (dotted line) and to room temperature (dashed line).}
\end{figure}

There is, however, a noticeable deviation for $V_{\textup{\tiny{SD}}}< -10$ mV between the simulated curve using the parabolic approximation for the potential barrier and the experimental data (Fig. 3). This indicates that the tunnel rate is substantially decreased below this value. Because the tunneling time in a single square barrier is given by $\tau_{\textup{\tiny{S}}} \sim \textup{sinh} \left(2 \alpha W_{\textup{\tiny{S}}} \right) $, where $\alpha$ is the attenuation constant, the potential barrier has to be substantially wider at energies lower than 5.3 meV (below the Fermi energy of the source lead when $V_{\textup{\tiny{SD}}}$=0). This effect is well confirmed by simulations if electron tunneling is significantly decreased for energy levels lying below -4$E_{\textup{\tiny{C}}}$. To investigate the origin of such barrier shape, and in particular, to distinguish between static (geometry) and dynamic (localization) effects, we proceeded with a series of thermal cycles (Fig. 4). 

After raising the device temperature to 150 K and slowly cooling it back to 300 mK, no NDC is observed. Instead, a plateau is present in the current-voltage characteristics. Following a different cooldown from room temperature, no NDC nor plateau could be observed. These results point towards the presence of trapped electrons that have been thermally activated. Further thermal cycles up to room temperature did not change the shape of the current voltage characteristics and only an inflexion remains (arrow in Fig. 4). This indicates that both a geometrical asymmetry in the barrier and electron trapping during cooldown are likely to be responsible for the NDC.

\subsubsection{Enhancement of localization}

\begin{figure}
\begin{center}
\includegraphics[width=85mm, bb=2 1 225 192]{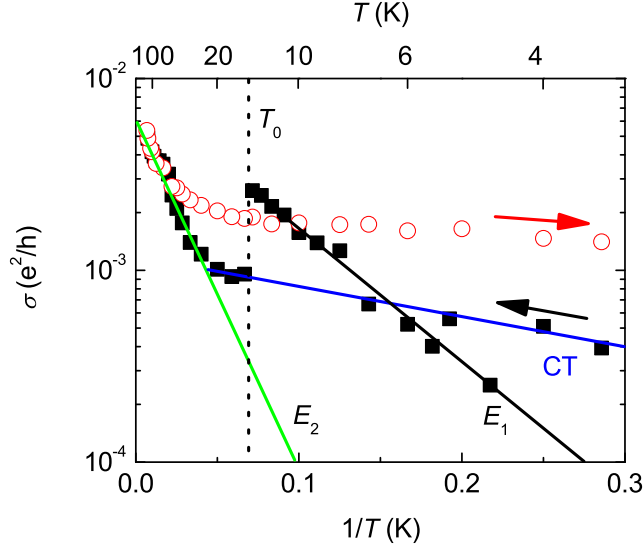}
\end{center}
\caption{\label{fig:figure5} $\sigma \left( T \right)$ with activated (solid lines) and cotunneling (CT) regimes when increasing and decreasing temperature from 3 to 150 K as shown by arrows. The dotted line shows the region of trap ionization associated with the electron delocalization temperature $T_0$. $V_{\textup{\tiny{SD}}} =$ 1.5 mV and $V_{\textup{\tiny{g}}}$ = -2.27 V.}
\end{figure}

Additional information on the trapping mechanism is obtained by analyzing the temperature dependence of the conductivity in the high temperature regime. To this end, we proceeded to a thermal cycle between 300 mK and 150 K and measured the variation in the height of a Coulomb peak in temperature (Fig. 5). The behavior is significantly different depending on whether the temperature is increased (squares) or decreased (circles). In the first case, the conductivity is limited by cotunneling at low temperature, as shown before (Fig. 2), then follows an activation behavior between 5 and 15 K, with an associated energy $E_1 = 1.39$ meV. The most noticeable feature is the abrupt decrease in conductivity at $T_0 \sim$ 15 K and the presence of a new activation process with an energy value $E_2 = 3.5$ meV above this temperature. In the second case, i. e. when the temperature is gradually decreased, the same activation process $E_2$ is present in the high temperature range but the process $E_1$ has disappeared. This suggests that $E_2$ is an intrinsic characteristic of the device, and so, corresponds to an activation of donors in silicon, whereas $E_1$ is associated with the way electrons are localized to impurity states during the cooldown. Such an irreversible and abrupt delocalization of traps is supported by the fact that $k_{B} T_0 \sim E_1$. For single donors in silicon or at low concentration, the ionization energy is expected to be close to 45 meV. This value, however, decreases substantially with the donor separation, and, at a concentration of $\sim 10^{19}$ cm$^{-3}$ similar to the one used in these devices, it is presumably close to zero \cite{Altermatt}. As a consequence, $E_2$ is likely to be associated with phosphorus donors at the Si/SiO$_2$ interface where localization effects may be more important. This increase of localization arises from the sensitivity of the ionization energy to the dielectric environment. Indeed, for degenerate silicon, we have \cite{Wacquez}:

\begin{eqnarray}\label{eqn:equation6}
E_2 \sim \frac{1}{4\pi \varepsilon_0} \frac{e^2}{2z} \frac{\varepsilon_{\textup{\tiny{Si}}}-\varepsilon_{\textup{\tiny{SiO}}_2}}{\varepsilon_{\textup{\tiny{Si}}} \left( \varepsilon_{\textup{\tiny{Si}}}+\varepsilon_{\textup{\tiny{SiO}}_2} \right) }
\end{eqnarray}

where $z$ is the distance to the SiO$_2$-Si interface and $\varepsilon_0$ the dielectric constant in free space.

Indeed for $z \sim 9$ nm, we obtain $E_2 \sim$ 3.5 meV.

Finally, we notice that the NDC disappears irreversibly above 17 K, thus linking the electron localization in the barrier to the presence of an NDC. Still, this process alone cannot explain the observed phenomenon as a whole. Instead a combination of structural defects in the barrier and localization enhancement due to electron trapping may explain the behaviors observed in temperature and gate voltage dependences. Structural defects in the barrier may arise from dislocations or the presence of impurities, silicon dangling bond defects or the formation of a phosphorus clusters that localize electrons that may form either during the implantation or the annealing stage.

After thermal cycling, the low temperature conductivity is higher, implying that the strength of localization has been decreased successfully.

\subsection {Low temperature and magnetic field dependences}

\subsubsection {Temperature effects}

\begin{figure}
\begin{center}
\includegraphics[width=85mm, bb=2 2 229 174]{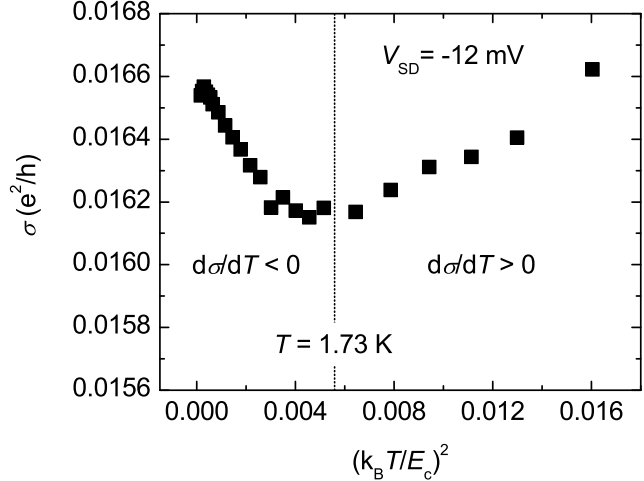}
\end{center}
\caption{\label{fig:figure6} Temperature dependence of the conductivity in the low temperature regime. A change in behavior is clearly noticeable at around 1.73 K. $E_{\textup{\tiny{C}}}$ is the charging energy of the quantum dot.}
\end{figure}

\begin{figure}
\begin{center}
\includegraphics[width=85mm, bb=2 2 250 164]{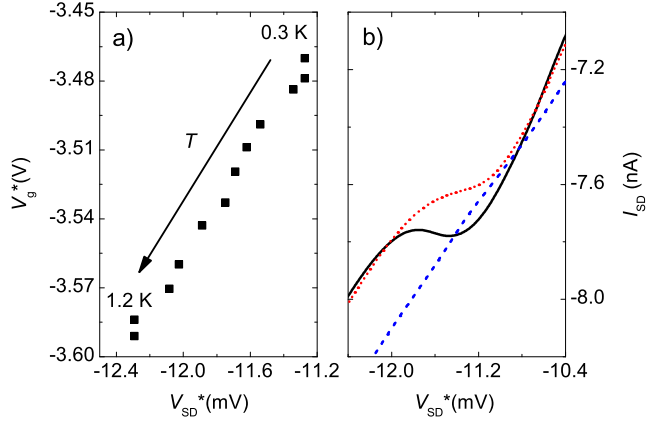}
\end{center}
\caption{\label{fig:figure7} (a) Variation of the position of the NDC in gate voltages with temperature. (b) Evolution of the shape and position of the NDC at 0.3 K (solid line), 1.5 K (dotted line) and 3 K (dashed line). The apparent shift towards positive $V_{\textup{\tiny{SD}}}$ is an effect of the displacement of the NDC in gate voltage, as described in the text.}
\end{figure}

Despite the high doping concentration, the silicon quantum dot shows an insulating behavior down to 300\,mK, for all gate voltages $V_{\textup{\tiny{g}}}$ i.e. the conductivity decreases with temperature (Figs. 2 and 5). However, for a small range of source-drain biases $V_{\textup{\tiny{SD}}}$ close in value to the the NDC position and in the absence of magnetic field, we observe an increase in conductivity $\sigma$ below $T_{\textup{\tiny{C}}}=$ 1.73\,K ($\sim 149\,\mu$eV) (Fig. 6). To understand the origin of the effect, we have measured the dependence of the conductivity on $V_{\textup{\tiny{SD}}}$ and $V_{\textup{\tiny{g}}}$, similarly to Figure 1, but at different temperatures up to 3 K . We find that the NDC position ($V_{\textup{\tiny{SD}}}^*$ and $V_{\textup{\tiny{g}}}^*$) is significantly affected by the temperature (Fig. 7a). Also, the NDC evolves into a plateau due to the presence of additional tunneling mechanisms above 1.5\,K, a value close to $T_{\textup{\tiny{C}}}$ (Fig. 7b). 

By raising the temperature of the device, the conductivity increases via cotunneling and the energy levels get broader, leading to a loss in the NDC visibility. The second effect of thermal broadening is a slight displacement of the NDC in source-drain bias of the order of $3.53\,k_{\textup{\tiny{B}}}T$ \cite{Beenaker}. This corresponds to the thermal broadening of a Coulomb peak in the classical and high temperature regime, so the expected shift at 1.2 K is about 830 $\mu$V, as measured (Fig. 7). The increase of temperature also modifies the tunneling rates via Eqs. 1 and 2, so that the fractional number of electrons in the quantum dot $n$ becomes temperature dependent and

\begin{eqnarray}\label{eqn:equation5}
n = \frac{\it{\Gamma}_{\textup{\tiny{D}}}}{\it{\Gamma}_{\textup{\tiny{D}}} + \it{\Gamma}_{\textup{\tiny{S}}}} \sim 1-\it{\gamma} \textup{exp} \left(\delta E / k_{\textup{\tiny{B}}}T \right)
\end{eqnarray}

where $\gamma$ is a constant depending on the device characteristics and $\delta E$ a function of $V_{\textup{\tiny{SD}}}$ and $V_{\textup{\tiny{g}}}$.

Consequently, Coulomb diamonds shift along $V_{\textup{\tiny{g}}}$ with temperature, so is the position of the NDC, as observed experimentally. Since Figure 6 is obtained at fixed values of $V_{\textup{\tiny{g}}}$ and $V_{\textup{\tiny{SD}}}$, the variation d$\sigma/$d$T$ reflects indeed the displacement of $V_{\textup{\tiny{SD}}}^*$ and $V_{\textup{\tiny{g}}}^*$ in the map of Figure 1, and consequently, the variation of the conductivity $\sigma \left( V_{\textup{\tiny{g}}}, V_{\textup{\tiny{SD}}} \right)$ at low temperature.

\subsubsection{Magnetic field effects}

\begin{figure}
\begin{center}
\includegraphics[width=85mm, bb=2 1 248 165]{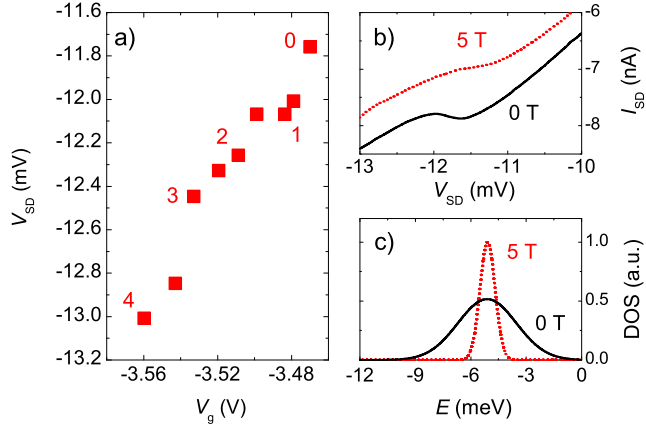}
\end{center}
\caption{\label{fig:figure8} a) NDC position at different magnetic fields (field number in red in inset) after thermal cycle. b) $I_{\textup{\tiny{SD}}} \left(V_{\textup{\tiny{SD}}}\right)$ at 0 (solid line) and 5 T (dotted line) before thermal cycle and at a fixed $V_{\textup{\tiny{g}}}$. The relative similar positions of the NDC is an effect of the dependence of the NDC position on $B$. c) Sketch of the effect of magnetic field on the density of states of the localized states located in the barrier.}
\end{figure}

Similarly to temperature effects, the application of a magnetic field $B$ perpendicular to the substrate modifies both the position of the NDC and its visibility (Fig. 8a). One may expect an increase in the localization strength at high $B$, and so, a more pronounced NDC. However, only an inflexion in the current voltage characteristic is visible (Fig. 8b). Unlike thermal cycles, this effect is reversible. The effect of the field does reduce the effective diameter of the quantum dot by localizing more strongly the electrons at the Si-SiO$_2$ interface. It also affects the capacitances values. However, this is the density of states (DOS) near the source barrier which is predominantly affected. The number of localized states in the barrier is conserved but the spread in energy level is reduced (band tails in the DOS disappearing) (Fig. 8c), so the NDC is smoothed. Such a modification of the DOS with $B$ also explains that the tunneling rates are locally increased when tunneling takes place. The effective electron number in the dot is then reduced and the NDC position in $V_{\textup{\tiny{g}}}$ is shifted accordingly. The combination between an increase in the charging energy and the disappearance of the DOS tails justifies the magnitude of the shift observed along $V_{\textup{\tiny{SD}}}$.

\section{Conclusions}

We have shown that the NDC that we observed could be explained by the presence of defects in one of the tunnel barrier and by an increase in the localization in that region due to electron trapping during the cooldown of the sample. Within this description, the change in the sign of d$\sigma/$d$T$ results from the combination of the presence of an NDC region at negative source-drain bias consecutive to localization effects and the displacement of its position in $V_{\textup{\tiny{g}}}$ and $V_{\textup{\tiny{SD}}}$ with temperature, an effect which is partly due to thermal broadening and a modification of the tunneling rates. The application of a perpendicular magnetic field mainly affects the DOS, leading to similar effects to the temperature, including a displacement of the NDC with respect to gate and source-drain biases and the disappearance of the NDC at high field. These effects may have implications on the electrostatic control over the tunnel barriers as well as the spin-dependent electron tunneling if such a method is used for reading out the spin states of an electron.

\section*{Acknowledgment}

This work was supported by Project for Developing Innovation Systems of the Ministry of Education, Culture, Sports, Science and Technology (MEXT), Japan and by Grants-in-Aid for Scientific Research from MEXT under Grant No. 22246040. T. Kodera would like to acknowledge JST-PRESTO for financial support. Two of the authors, T. F. and A. R., contributed equally to the work.


\section*{References}

\begin{thebibliography}{23}

\bibitem{Anderson} P. W. Anderson, Phys. Rev. 109, 5, 1492 (1958)
\bibitem{Davies} R. A. Davies, C. C. Dean and M. Pepper, Surf. Sci., 142, 25 (1984); D. J. Thouless, Physica B, 109–110, 3, 1523 (1982)
\bibitem{Abrahams} A. Miller and E. Abrahams, Phys. Rev. 120, 745 (1960); N. F. Mott, Phil. Mag. 19, 835 (1969)
\bibitem{Ferrus1} T. Ferrus, R. George, C. H. W. Barnes, and M. Pepper, Appl. Phys. Lett. 97, 14, 142108 (2010)
\bibitem{MIT} N. F. Mott, Rev. Mod. Phys. 40, 677 (1968); \textit{ibid} Phil. Mag. 20, 163, 1 (1969); \textit{ibid} Phil. Mag. 26, 1015 (1972)
\bibitem{Zhitenev} N. B. Zhitenev, M. Brodsky, R. C. Ashoori, L. N. Pfeiffer and K. W. West, Science 285, 5428, 715 (1999)
\bibitem{Ferrus2} T. Ferrus, A. Rossi, M. Tanner, G. Podd, P. Chapman, and D. A. Williams, New J. Phys., 13, 10, 103012 (2011)
\bibitem{Sanquer} M. Hofheinz, X. Jehl, M. Sanquer, G. Molas, M. Vinet, S. Deleonibus, Euro. Phys. Jour. B. 54, 299-307 (2006)
\bibitem{Sanquer2} V. N. Golovach, X. Jehl, M. Houzet, M. Pierre, B. Roche, M. Sanquer, L. I. Glazman, Phys. Rev. B 83, 075401 (2011)
\bibitem{Rogge} H. Sellier, G. P. Lansbergen, J. Caro, N. Collaert, I. Ferain, M. Jurczak, S. Biesemans, S. Rogge,	Phys. Rev. Lett. 97, 206805 (2006)
\bibitem{spin blockade} H. W. Liu, T. Fujisawa, Y. Ono, H. Inokawa, A. Fujiwara, K. Takashina, and Y. Hirayama, Phys. Rev. B 77, 073310 (2008)
\bibitem{spin qubit} F. H. L. Koppens, C. Buizert, K. J. Tielrooij, I. T. Vink, K. C. Nowack, T. Meunier, L. P. Kouwenhoven, and L. M. K. Vandersypen, Nature 442, 766 (2006)
\bibitem{Pierre} M. Pierre, M. Hofheinz, X. Jehl, M. Sanquer, G. Molas, M. Vinet, and S. Deleonibus, Eur. Phys. J. B 70, 475–481 (2009)
\bibitem{Hitachi} K. Hitachi, A. Inoue, A. Oiwa, M. Yamamoto, M. Pioro-Ladriere, Y. Tokura and S. Tarucha, J. Phys: Conf. Ser. 150, 022026 (2009)
\bibitem{Rossi} A. Rossi, T. Ferrus, W. Lin, T. Kodera, D. A. Williams, and S. Oda, Appl. Phys. Lett. 98, 133506 (2011)
\bibitem{Rates} Tunneling rates are generally defined for single tunneling events from the Fermi energy of the source lead to a well defined energy level in the dot. The value of this rate can be measured directly by the use of single shot measurement techniques. However, in the case of standard DC measurements where the integration time is much larger than the tunneling time, many tunneling events happen. In doped material, the disorder lifts the level degeneracy and the measured tunneling rate across the barrier is modified by including a factor $\gamma \sim N^2$ where $N$ is the number of active dopants on each side of the barrier. By taking into account a depletion region of $\sim$ 15 nm at the Si-SiO$_2$ interface and an effective SOI thickness of 20 nm after oxidation, we find $N \sim$ 392 atoms and $\gamma \sim 6.17 \times 10^5$, if spin degeneracy is taken into account.
\bibitem{Furusaki} A. Furusaki and K. A. Matveev, Phys. Rev. B 53, 16676 (1995)
\bibitem{Glatti} D. C. Glattli, C. Pasquier, U. Meirav, F. I. B. Williams, Y. Jin, and B. Etienne, Z. Phys. B - Condens. Matter 85, 375 (1991)
\bibitem{MacLean} K. MacLean, S. Amasha, J. P. Radu, D. M. Zumb\"{u}hl, M. A. Kastner, M. P. Hanson, and A. C. Gossard, Phys. Rev. Lett. 98, 036802 (2007)
\bibitem{ElecTemp} The power of the heat transfer is $\propto {T_e}^{p+2}-{T_0}^{p+2}$ where $p$ is related to the temperature dependence of electron-phonon scattering time $\tau^{-1} \propto T^p$. In a metal, $p=3$ and, if the electronic system is decoupled from the phonon bath, it is expected that $p=0$. In silicon, scattering by acoustic phonons is the dominant process at low temperature and $p=3/2$. In the cotunneling regime, the power of the Joule effect is $\sim {V_{\textup{\tiny{SD}}}}^4$, leading to a bias dependent electron temperature $T_e \sim {V_{\textup{\tiny{SD}}}}^{8/7}$ at high source drain biases.
\bibitem{Altermatt} P. P. Altermatt, A. Schenk, B. Schmith\"{u}sen, and G. Heiser, J. Appl. Phys. 100, 113714 (2006); \textit{ibid} J. Appl. Phys. 100, 113715 (2006) and references within.
\bibitem{Wacquez} M. Pierre, R. Wacquez, X. Jehl, M. Sanquer, M. Vinet and O. Cueto, Nature Nanotechnology 5, 133 (2010)
\bibitem{Beenaker} C. W. J. Beenakker, Phys. Rev. B 44, 1646 (1991)

\end{thebibliography}
\end{document}